\documentclass[12pt]{article}
\usepackage{graphicx}
\usepackage{epsfig}
\usepackage{amsmath}

\newcommand{\spone}{0.9}  

\newcommand{\sptwo}{1.4}
\newcommand{\spthree}{2.4}

\newcommand{\singlespace}{\edef\baselinestretch{\spone}\Large\normalsize}

\newcommand{\doublespace}{\edef\baselinestretch{\sptwo}\Large\normalsize}
\newcommand{\threespace}{\edef\baselinestretch{\spthree}\Large\normalsize}

\singlespace
\threespace
\doublespace

\everymath={\displaystyle}
\begin{document}
\doublespace

\begin{center}
{\bf {\large Nonlinear Realization of N=2 Superconformal \\
Symmetry and Brane Effective Actions
}\\
$~$\\
Lu-Xin Liu\\
{\it Department of Physics, Purdue University,\\ 
West Lafayette, IN 47907, USA\\
liul@physics.purdue.edu \\}}
\end{center}
\vspace{25pt}
{\bf Abstract.} 
Due to the incompatibility of the nonlinear realization of superconformal symmetry and dilatation 
symmetry with the dilaton as the compensator field, in the present paper it shows an alternative 
mechanism of spontaneous breaking the N=2 superconformal symmetry to the N=0 case. 
By using the approach of nonlinear transformations it is found that it leads to a space-filling 
brane theory with Weyl scale W(1,3) symmetry. The dynamics of the resulting Weyl scale
 invariant brane, along with that of other Nambu-Goldstone fields, is derived in terms of the 
building blocks of the vierbein and the covariant derivative from the Maurer-Cartan one-forms.
 A general coupling of the matter fields localized on the brane world volume to these NG fields is also constructed.

\vspace{60pt}

\pagebreak

\begin{flushleft}
{\large I. Introduction}
\end{flushleft} 
\vspace{5pt}

In 1960s, being a powerful tool nonlinear realization method was used to study the 
low energy dynamics of the chiral symmetry [1], especially for the case when the full 
symmetry was spontaneously broken and the partners of the Goldstone bosons became massive 
and decoupled from the dynamics of the NG bosons. Later on, in Ref.[2] a general approach 
of nonlinear realization was given in terms of compact semisimple Lie groups, in which the 
full symmetry group was realized in terms of the nonlinear transformations of the NG fields 
which were promoted from the exponential parameters of the Coset representative elements, 
and these transformations were isomorphic to the left action of the general group elements 
on the Coset. Therefore, the transformations of the NG fields, together with that of the 
spectator fields which transform linearly under the unbroken subgroup $H$, give a complete 
representation of the full symmetry group. The resulting phenomenological Lagrangian, which is 
model independent, becomes an effective theory at energies far below the scale of spontaneous symmetry breaking. 
Nonlinear realization was also extended to include spacetime symmetries [3, 4], in which the spacetime translational 
generators and the broken symmetry generators transform independently under the stability group. Recently, extensive 
research work of nonlinear realization has been extended to describe the dynamics of the brane theories [5, 6, 7]. 
Consider a topological defect embedded in a target space. Its world volume then has the symmetries of the unbroken 
stability subgroup, whileas its long wave oscillations into the co-dimensional (super) space are described by the 
Nambu-Goldstone (Goldstino) modes associated with those broken symmetries [6, 7].

     In this paper, we construct the nonlinear realization of spontaneously broken N=2 superconformal 
symmetry by using Weyl scale invariant brane theories. As for the extended SUSY theory which includes more
 than one spinor supercharge, it is possible that it may arise from higher dimensional theories in some supersymmetric
 models or in some effective theories derived from higher dimensions via dimensional reductions. It is also well known 
that it plays an important role in understanding nonperturbative aspects of supersymmetric theories [8]. In the present context, 
it has been discussed that the N=2 superconformal symmetry could be spontaneously broken by different approaches [9]. 
It may be partially broken down to the N=1 Super Poincare symmetry or to N=2 Super Poincare symmetry. Since supersymmetry 
must be broken for a realistic theory, in the present paper we consider the case when N=2 superconformal symmetry is totally broken 
down to N=0 supersymmetry.

         It has been pointed out that the nonlinear realizations of supersymmetry and dilatation symmetry are not 
compatible when the dilaton is taken as the compensator field [10]. Therefore, in the present paper, one 
of its purposes is to show an alternative mechanism that describes the total supersymmetry breaking of the 
N=2 superconformal group. As it shows below, it would lead to a theory with Weyl scale invariant symmetry.

     Consider a model, a space filling brane embedded in D=4 spacetime. Taking the 
static gauge, the spacetime coordinates $x^\mu$ therefore line up with the variables $\xi ^\mu$ which 
parameterize the brane world volume, i.e. $ \xi ^\mu   = x^\mu$ (we take static gauge in what follows of 
this paper). The target superspace has the N=2 superconformal symmetry, and the 
embedded submanifold has unbroken Weyl $W(1,3)$ symmetry, which is formed by the set $\{ P_\mu  ,D,M_{\mu \nu } \}$. Then 
the low energy effective action of the system, which is scale invariant under the 
transformation of $\xi ^\mu   \to e^d \xi ^\mu $, can be described by long wave oscillations of the brane into the 
superspace associated to the Grassmann coordinate directions $\lambda ,\bar \lambda$, as well as the dynamics 
of the Nambu-Goldstone mode corresponding to broken generator $A$ of the internal space.

        On the other hand, the purpose of the paper is to explore more features of the well known $AdS/CFT$ correspondence [11, 12]. 
In the spirit of the $AdS/CFT$ correspondence, following the outline sketched by this paper one can find it paves a way to 
embed a probe brane into a $AdS \times S$ background and realize the supersymmetric isometry of the target space. Associated 
with the brane, it is expected that there would be no destabilized terms due to the brane's oscillations into the transverse 
spatial directions. We hope that would shed some 
light on the appealing aspects of the $AdS/CFT$ correspondence in terms of the brane world scenarios.

               The paper is organized as follows. First, in section II, we introduce the N=2 superconformal algebra, 
and construct the Coset structure in terms of the unbroken $W(1,3)$ subgroup. The general infinitesimal 
transformations of the coordinates, as well as that of the $Q_{\alpha i}$ type Goldstinos, are introduced through the 
action of the full group elements on the Coset representatives, whileas in section III the $S^{\alpha i}$ type Goldstino 
spinors are proved to be superfluous and are eliminated by imposing covariant constraints. Also in that section, 
it follows that the vierbein and the covariant derivative of the NG field can be derived by means of the Maurer-Cartan 
one-forms. The effective scale invariant brane action is then constructed in terms of these building blocks. In section IV, 
the general coupling of the matter fields localized on the brane world volume to these NG modes is introduced. The infeasibility 
of taking the Lorentz group as the stability group for the spontaneous breaking of the full group is also pointed out, which is due 
to the fact that the nonlinear realization of the N=2 superconformal symmetry and the dilatation symmetry are not compatible 
when considering the dilaton field as the compensator field.

\vspace{5pt}
\begin{flushleft}
{\large II. Nonlinear Realization of the N=2 Superconformal Symmetry}
\end{flushleft}
\vspace{5pt}

The N=2 Superconformal algebra is isomorphic to the simple Lie superalgebra $su(2,2|2)$, 
real form of $sl(4|2)$. Its algebra includes the conformal algebra:
$$ [M_{\mu \nu } ,M_{\rho \sigma } ] = i(\eta _{\mu \sigma } M_{\nu \rho }  + \eta _{\nu \rho } M_{\mu \sigma }  - \eta _{\mu \rho } M_{\nu \sigma }  - \eta _{\nu \sigma } M_{\mu \rho } ) \\$$
$$[P_\mu  ,M_{\kappa \lambda } ] = i(\eta _{\mu \kappa } P_\lambda   - \eta _{\mu \lambda } P_\kappa  ), [K_\mu  ,M_{\kappa \lambda } ] = i(\eta _{\mu \kappa } K_\lambda   - \eta _{\mu \lambda } K_\kappa  ) \\$$
$$[P_\mu  ,D] = iP_\mu,   
 [K_\mu  ,D] =  - iK_\mu,   
 [M_{\mu \nu } ,D] = 0 \\$$
$$[P_\mu  ,K_\nu  ] = 2i(\eta _{\mu \nu } D - M_{\mu \nu } ) \\  \eqno{(1)}
$$
in which $\eta ^{\mu \nu }  = ( + , - , - , - , - )$. It also has two different types of spinor charges $Q_{\alpha i}$ and $\bar S_i^{\dot \alpha }$, and the 
commutation relations of these fermion-type charges with the conformal 
group generators and the internal group $SU(2) \times U(1)_R$ generators have the form
$$[Q_{\alpha i} ,K_\mu  ] = \sigma _{\mu \alpha \dot \beta }^{} \bar S_i^{\dot \beta }, 
 [\bar S_i^{\dot \alpha } ,K_\mu  ] = 0, 
 [\bar S_i^{\dot \alpha } ,P_\mu  ] = \bar \sigma _\mu  ^{\dot \alpha \beta } Q_{\beta i}  \\$$
$$[Q_{\alpha i} ,D] = \frac{1}{2}iQ_{\alpha i}, 
 [Q_{\alpha i} ,A] = \frac{1}{2}Q_{\alpha i}, 
 [\bar S_i^{\dot \alpha } ,D] =  - \frac{1}{2}i\bar S_i^{\dot \alpha } , 
 [S^{\alpha i} ,A] =  - \frac{1}{2}S^{\alpha i}  \\  $$
$$[Q_\alpha  ,M_{\mu \nu } ] = \frac{1}{2}(\sigma _{\mu \nu } )_\alpha  ^{\\\ \beta}  Q_\beta , 
 [\bar Q^{\dot \alpha } ,M_{\mu \nu } ] = \frac{1}{2}(\bar \sigma _{\mu \nu } )^{\dot \alpha } _{\\\ \dot \beta } \bar Q^{\dot \beta }  \\ $$
$$ [Q_{\alpha i} ,T_j ^k ] = \delta _i^k Q_{\alpha j}  - \frac{1}{2}\delta _j^k Q_{\alpha i}, 
 [\bar S_i^{\dot \alpha } ,T_j ^k ] = \delta _i^k \bar S_j^{\dot \alpha }  - \frac{1}{2}\delta _j^k \bar S_i^{\dot \alpha }  \\ 
\eqno{(2)}$$
When $N>1$, there is no central charge, and the (anti)commutation relations among the fermion-type charges are
$$ \{ Q_{\alpha i} ,\bar Q_{\dot \alpha } ^j \}  = 2\delta _i^j \sigma _{\alpha \dot \alpha }^\mu  P_\mu, \{ \bar S_i^{\dot \alpha } ,S^{\beta j} \}  = 2\delta _i^j \bar \sigma ^{\mu \dot \alpha \beta } K_\mu   \\ $$
$$\{ Q_{\alpha i} ,Q_{\beta j} \}  = 0, 
 \{ Q_{\alpha i} ,\bar S_j^{\dot \beta } \}  = 0, 
 \{ \bar S_i^{\dot \alpha } ,\bar S_j^{\dot \beta } \}  = 0 \\ $$
$$ \{ Q_{\alpha i} ,S^{\beta j} \}  = 
\delta _i^j [(\sigma _{}^{\mu \nu } )_\alpha ^{\\\ \beta}  M_{\mu \nu }  - 
2iD\delta _\alpha ^{\\\ \beta}  ] - 4\delta _\alpha ^{\\\ \beta}  (T_i^j  + \frac{1}{2}\delta _i^j A); 
\\ \eqno{(3)}$$
we take the notation $ \varepsilon _{\alpha \beta }  = \varepsilon _{\dot \alpha \dot \beta }  =  - \varepsilon ^{\alpha \beta }  =  - \varepsilon ^{\dot \alpha \dot \beta }   = \left( {\begin{array}{*{20}c}
   0 & 1  \\
   { - 1} & 0  \\
\end{array}} \right) $, and $\sigma _{\mu \nu }  = \frac{1}{2}i(\sigma _\mu  \bar \sigma _\nu   - \sigma _\nu  \bar \sigma _\mu  )  $. Where $A$ is the generator of $U(1)_R $, and the $SU(2)$ generators $T_i^j $ satisfy
$$[T_i^j ,T_k ^l ] = \delta _i^l T_k^j  - \delta _k^j T_i^l \\ \eqno{(4)}$$                                                                                                    
where $i,j = 1,2.$ Consider the case when the N=2 superconformal symmetry is spontaneously broken down to the N=0 case. 
We choose the unbroken subgroup as 
$$W(1,3) \times SU(2) \\  \eqno{(5)} $$                                                                                                    
in which $ W(1,3)$ is the Weyl group, formed by the set $\{ P_\mu  ,D,M_{\mu \nu } \} $[10]. Therefore, the group elements of the stability 
subgroup $H$ are written as 
$$ h = e^{i(fD + m^{\mu \nu } M_{\mu \nu }  + t_j ^i T_i^j )}  \\ \eqno{(6)} $$                                                                                                          which is spanned by the set of generators $\{ D,M_{\mu \nu } ,T_i^j \} $, and from Eqs.(1, 2) we can find these generators are 
automorphism of the broken generators $ Q,\bar Q,S,\bar S,K^\mu  ,A $ which are associated 
with the collective coordinates $\lambda ,\bar \lambda ,\varsigma ,\bar \varsigma ,\phi ^\mu $ and $a$ respectively. Therefore, 
the Coset is shown to be 
$$ \Omega  = G/H \\ \eqno{(7)} $$                                                                                                                                                                                                                         
In static gauge their representative elements can be parameterized as                                                    
$$\Omega  = e^{ix^\mu  P_\mu  } e^{i[\lambda Q + \bar \lambda \bar Q]} e^{i(\varsigma S + \bar \varsigma \bar S)} e^{i\phi ^\mu  K_\mu  } e^{iaA}  \\ \eqno{(8)} $$                                                                                  The left 
action of the general infinitesimal elements of the full group $G$
$$ g = e^{i(a^\mu  P_\mu   + qQ + \bar q\bar Q + sS + \bar s\bar S + b^\mu  K_\mu   + rA + \rho ^{\mu \nu } M_{\mu \nu }  + dD + \varepsilon _j ^i T_i^j )}  \\ \eqno{(9)} $$                                                                            
on the Coset representative elements of Eq. (8) can be uniquely decomposed as the product of the 
Coset $\Omega '$ and the stability group elements of $H$, i.e.                                                                               
$$g\Omega  = \Omega 'h \\ 
\eqno{(10)} $$                                                                                                                      
Explicitly, we have
\begin{align*}
&e^{i(a^\mu  P_\mu   + qQ + \bar q\bar Q + sS + \bar s\bar S + b^\mu  K_\mu   + rA + \rho ^{\mu \nu } M_{\mu \nu }  + dD + \varepsilon _j ^i T_i^j )}  
 e^{ix^\mu  P_\mu  } e^{i[\lambda Q + \bar \lambda \bar Q]} e^{i(\varsigma S + \bar \varsigma \bar S)} e^{i\phi ^\mu  K_\mu  } e^{iaA} \\
= &e^{ix'^\mu  P_\mu  } e^{i[\lambda 'Q + \bar \lambda '\bar Q]} e^{i(\varsigma 'S + \bar \varsigma '\bar S)} e^{i\phi '^\mu  K_\mu  } e^{ia'A}   
 e^{i(fD + m^{\mu \nu } M_{\mu \nu }  + t_j ^i T_i^j )} 
\tag{11}
\end{align*}
The infinitesimal transformation of the preferred fields can be derived up to the first order by 
considering the variation $ \delta g $ of the group elements $g$ with
$$\delta g \\ 
  = i(a^\mu  p_\mu   + qQ + \bar q\bar Q + sS + \bar s\bar S + b^\mu  K_\mu   + rA + \rho ^{\mu \nu } M_{\mu \nu }  + dD + \varepsilon _j ^i T_i^j ) \\   \eqno{(12)} $$                    
Hence, Eq.(10) becomes
$$(1 + \delta g)\Omega  = (\Omega  + \delta \Omega )(1 + \delta h) \\ \eqno{(13)} $$                                                                                          
furthermore
 $$\Omega ^{ - 1} \delta g\Omega  - \Omega ^{ - 1} \delta \Omega  = \delta h \\ \eqno{(14)} $$                                                                                             
Then it is found 
\begin{align*}
& \Omega ^{ - 1} i(a^\mu  p_\mu   + qQ + \bar q\bar Q + sS + \bar s\bar S + b^\mu  K_\mu   + rA +  \\
& \rho ^{\mu \nu } M_{\mu \nu }  + dD + \varepsilon _j ^i T_i^j )\Omega  - \Omega ^{ - 1} \delta \Omega  \\                                  
= & i(fD + m^{\mu \nu } M_{\mu \nu }  + t_j ^i T_i^j ) 
\tag{15}
\end{align*}                                         
Consider the pure shift induced by Poincare translation in the four dimensional 
spacetime, i.e., taking $g = e^{ia^\mu  P_\mu  }$, it is found 
$$\delta x^\mu   = a^\mu, 
 \delta \lambda  = \delta \varsigma  = \delta \phi ^\mu   = \delta a = 0 \\ \eqno{(16)} $$                                                                                
and 
$$f = m^{\mu \nu }  = t_j^i  = 0. \\ \eqno{(17)}  $$                                                       Also, as for the pure shift $g = e^{i(qQ + \bar q\bar Q)} $ in the superspace, we have                                                 
$$ \delta x^\mu   =  - i(\lambda ^i \sigma ^\mu  \bar q_i  - q^i \sigma ^\mu  \bar \lambda _i ); \\ $$                                                        
$$ \delta \lambda ^i  = q^i;   
 \delta \bar \lambda _i  = \bar q_i;  \\ \eqno{(18)} $$ 
as well as $\delta \varsigma  = \delta \phi ^\mu   = \delta a =   
 f = m^{\mu \nu }  = t_j^i  = 0  $. Here $\delta \lambda ^i$ etc are the total variation of the fields, i.e. $\delta \lambda ^i  = \lambda '(x') - \lambda (x) $, and the intrinsic variation of 
the Goldstino fields $\lambda ^i  $ is given by
$$\delta_{in} \lambda ^i  = \lambda '(x') - \lambda (x') ={\delta \lambda ^i}-(\lambda (x') - \lambda (x)) \\ $$                                                
$$= q^i  + i({\lambda^j \sigma ^\mu  {\bar q}_j}-q^j\sigma ^\mu  {\bar \lambda}_j )\partial _\mu  \lambda ^i (x) \\ 
\eqno{(19)}  $$                                                                                    
which is just the Akulov-Volkov nonlinear transformation of the Goldstino fields [14, 15] for the 
extended Supersymmetries. Besides, by comparing the coefficients of the $Q_i$ and $P_\mu$ from both sides of Eq.(15) 
the general nonlinear transformations of the coordinates and the associated Goldstino fields can be 
derived as following (the Goldstino fields $\varsigma _i  $ corresponding to $S^i $ are superfluous and can be written as 
functions of $ \bar \lambda _i (x) $,see Eq.(32')):
\begin{align*}
\delta x^\mu   &= a^\mu  
  - i2(\lambda ^i \sigma ^\mu  \bar q_i  - q^i \sigma ^\mu  \bar \lambda _i )  
  + i(\lambda ^i \sigma ^\mu  d\bar \lambda _i  - d\lambda ^i \sigma ^\mu  \bar \lambda _i ) 
  + r\lambda ^i \sigma ^\mu  \bar \lambda _i  \\ 
  &- \frac{1}{{3!}}i(2\lambda ^i \sigma ^\mu  B_{4i}  - 2A_4^i \sigma ^\mu  \bar \lambda _i ) 
  - \frac{1}{{3!}}i(2\lambda ^i \sigma ^\mu  B_{5i}  - 2A_5^i \sigma ^\mu  \bar \lambda _i ) \\ 
  &- 2\rho ^{\mu \nu } x_\nu   + (x^{\mu '} b^{\nu '}  - \frac{1}{2}\rho ^{\mu '\nu '} )(\lambda ^i \sigma ^\mu  \bar \lambda _i \bar \sigma _{\mu '\nu '}  - \lambda ^i \sigma _{\mu '\nu '} \sigma ^\mu  \bar \lambda _i ) -  \\
 &\frac{1}{{4!}} \cdot 2i(\lambda ^i \sigma ^\mu  B_{6i}  - A_6^i \sigma ^\mu  \bar \lambda _i ) +  
 2x^\nu  b_\nu  x^\mu   - x^\nu  x_\nu  b^\mu   
  + d \cdot x^\mu   \\ 
  &+ \varepsilon _k^j (2\lambda ^k \sigma ^\mu  \bar \lambda _j  - \lambda ^i \sigma ^\mu  \delta _j^k \bar \lambda _i )
  + 2\lambda ^i \sigma ^\nu  x_\nu  s_i \sigma ^\mu   - 2x^\nu  \bar s^i \bar \sigma _\nu  \sigma ^\mu  \bar \lambda _i  
\tag{20}
\end{align*}
and
\begin{align*}
\delta \lambda ^i & = q^i  - i\frac{1}{2}r\lambda ^i  + ix^\mu  \bar s^i \bar \sigma _\mu   + \frac{1}{2}A_5^i  + \frac{1}{2}A_4^i  - \frac{1}{2}i\rho ^{\mu \nu } \lambda ^i \sigma _{\mu \nu }  + \frac{1}{{3!}}A_6^i  + \frac{1}{2}d \cdot \lambda ^i  \\ 
& + x^\mu  b_\mu  \lambda ^i  + ix^\mu  b^\nu  \lambda ^i \sigma _{\mu \nu }  - i\varepsilon _k^j (\lambda ^k \delta _j^i  - \frac{1}{2}\delta _j^k \lambda ^i ); \\
\delta \bar \lambda _i  &= \bar q_i  + i\frac{1}{2}r\bar \lambda _i  + ix^\mu  s_i \sigma ^\mu   + \frac{1}{2}B_{5i}  + \frac{1}{2}B_{4i}  - \frac{1}{2}i\rho ^{\mu \nu } \bar \lambda _i \bar \sigma _{\mu \nu }  + \frac{1}{{3!}}B_{6i}  + \frac{1}{2}d \cdot \bar \lambda _i  \\ 
 & + x^\mu  b_\mu  \bar \lambda _i  + ix^\mu  b^\nu  \bar \lambda _i \bar \sigma _{\mu \nu }  + i\varepsilon _k^j (\bar \lambda _j \delta _i^k  - \frac{1}{2}\delta _j^k \bar \lambda _i ) 
\tag{21}
\end{align*}
where the inhomogeneous terms play important roles in the nonlinear transformation and signal 
the spontaneous breaking of the extended supersymmetries. (see Appendix for definitions of the $A$ s 
and $B$ s in these equations) 

\vspace{5pt}
\begin{flushleft}
{\large III. The Effective Action}
\end{flushleft}
\vspace{5pt}

   The effective action can be derived by using the Maurer-Cartan one-forms which are 
expanded with respect to the $su(2,2|2)$ generators:             
\begin{align*}
\Omega ^{ - 1} d\Omega  &=i(\omega ^a P_a  + \omega _Q^k Q_k  + \bar \omega _{\bar Qk} \bar Q^k  + 
\omega _{Sk} S^k  + \bar \omega _{\hat S}^k \bar S_k  + \omega _k^a K_a  + \omega _A A \\ 
& + \omega _D D + \omega _Tj^i T_i^j  + \omega _M^{ab} M_{ab} )
\tag{22}
\end{align*}
Explicitly, they have the form
\begin{align*}
i\omega ^a &=   
 idx^a  + \lambda ^i \sigma ^\mu  d\bar \lambda _i  - d\lambda ^i \sigma ^\mu  \bar \lambda _i;  \\ 
 i\omega _Q^k  &=   
 dx^\mu  \bar \varsigma ^k \bar \sigma _\mu  e^{\frac{1}{2}ia}  + id\lambda ^k e^{\frac{1}{2}ia}  - i(\lambda ^i \sigma ^\mu  d\bar \lambda _i  - d\lambda ^i \sigma ^\mu  \bar \lambda _i )\bar \varsigma ^k \bar \sigma _\mu  e^{\frac{1}{2}ia};  \\ 
 i\bar \omega _{\bar Qk}  &=   
 dx^\mu  \varsigma _k \sigma _\mu  e^{ - \frac{1}{2}ia}  + id\bar \lambda _k e^{ - \frac{1}{2}ia}  - i(\lambda ^i \sigma ^\mu  d\bar \lambda _i  - d\lambda ^i \sigma ^\mu  \bar \lambda _i )\varsigma _k \sigma _\mu  e^{ - \frac{1}{2}ia};  \\ 
 i\omega _{Sk}  &=   
 (i\phi ^\mu  dx^\nu  \varsigma _k \sigma _\nu  \bar \sigma _\mu   + \frac{1}{{3!}}\left. {A_{1k} } \right|_{a^\mu   = idx^\mu  } )e^{ - i\frac{a}{2}}  + [\frac{1}{2}\left. {A_{2k} } \right|_{q_i  = id\lambda _i }  + \frac{1}{2}A_3 \left. {A_{3k} } \right|_{\bar q_i  = id\bar \lambda _i }  +  \\ 
 &\frac{1}{{3!}}\left. {A_{1k} } \right|_{a^\mu   = \lambda \sigma ^\mu  d\bar \lambda  - d\lambda \sigma ^\mu  \bar \lambda }  + i\phi ^\mu  (id\bar \lambda _k  -   
 i(\lambda ^i \sigma ^\nu  d\bar \lambda _i  - d\lambda ^i \sigma ^\nu  \bar \lambda _i )\varsigma _k \sigma _\nu  )\bar \sigma _\mu  ]e^{ - i\frac{a}{2}}  + id\varsigma _k e^{ - i\frac{a}{2}}  \\ 
 i\bar \omega _{\bar S}^k & =   
 (i\phi ^\mu  dx^\nu  \bar \varsigma ^k \bar \sigma _\nu  \sigma _\mu   + \frac{1}{{3!}}\left. {B_1^k } \right|_{a^\mu   = idx^\mu  } )e^{i\frac{a}{2}}  + [\frac{1}{2}\left. {B_2^k } \right|_{q^i  = id\lambda _i }  + \frac{1}{2}\left. {B_3^k } \right|_{\bar q_i  = id\lambda _i }  +  \\ 
 &\frac{1}{{3!}}\left. {B_1^k } \right|_{a^\varpi   = \lambda \sigma ^\mu  d\bar \lambda  - d\lambda \sigma ^\mu  \bar \lambda }  - \phi ^\mu  (d\bar \lambda ^k  - (\lambda ^i \sigma ^\nu  d\bar \lambda _i  - d\lambda ^i \sigma ^\nu  \bar \lambda _i )\bar \varsigma ^k \bar \sigma _\nu  )\sigma _\mu  ]e^{i\frac{a}{2}}  + id\bar \varsigma ^k e^{i\frac{a}{2}}  \\ 
 i\omega _k^a  &=   
  - i\phi ^{\mu '} \phi _{\mu '} dx^a  - (\bar \varsigma ^i \bar \sigma _\mu  \varsigma _i  + \bar \varsigma ^i \varsigma _i \sigma _\mu  )dx^\mu  \phi ^a   
  + \frac{1}{{4!}}2i(\left. {B_1^k } \right|_{a^\mu   = idx^\mu  } \bar \sigma ^a \varsigma _k  -  \\ 
 & \bar \varsigma ^k \bar \sigma ^a \left. {A_{1k}^{} } \right|_{a^\mu   = idx^\mu  } ) - i\bar \varsigma ^i (\bar \sigma _{\mu '} \sigma ^{a\nu } \varsigma _i  - \bar \sigma ^{a\nu } \bar \varsigma _i \sigma _{\mu '} )dx^{\mu '} \phi _\nu   + X^a  \\ 
 & - d\bar \varsigma ^i \bar \sigma ^a \varsigma _i  + \bar \varsigma ^i \bar \sigma ^a d\varsigma _i  + id\phi ^a  ;\\ 
 i\omega _A  &=  
 ida + 2d\lambda ^i \varsigma _i  - 2\bar \varsigma ^i d\bar \lambda _i  - \bar \varsigma ^i \bar \sigma _\mu  (\lambda ^j \sigma ^\mu  d\bar \lambda _j  - d\lambda ^j \sigma ^\mu  \bar \lambda _j )\varsigma _i  \\ 
  &+ \bar \varsigma ^i \varsigma _i \sigma _\mu  (\lambda ^j \sigma ^\mu  d\bar \lambda _j  - d\lambda ^j \sigma ^\mu  \bar \lambda _j ) + i( - \bar \varsigma ^i \bar \sigma _\mu  \varsigma _i  + \bar \varsigma ^i \varsigma _i \sigma _\mu  )dx^\mu;   \\ 
 i\omega _D  &=  
  - 2i\phi ^\mu  dx_\mu   + (\bar \varsigma ^i \bar \sigma _\mu  \varsigma _i  + \bar \varsigma ^i \varsigma _i \sigma _\mu  )dx^\mu   + 2id\lambda ^i \varsigma _i  + 2i\bar \varsigma ^i d\bar \lambda _i  \\ 
 & - i\bar \varsigma ^i \varsigma _i \sigma _\mu  (\lambda ^j \sigma ^\mu  d\bar \lambda _j  - d\lambda ^j \sigma ^\mu  \bar \lambda _j ) - 2\varphi ^\mu  (\lambda ^j \sigma _\mu  d\bar \lambda _j  - d\lambda ^j \sigma _\mu  \bar \lambda _j ) \\ 
  &- i\bar \varsigma ^i \bar \sigma _\mu  (\lambda ^j \sigma ^\mu  d\bar \lambda _j  - d\lambda ^j \sigma ^\mu  \bar \lambda _j )\varsigma _i ; \\ 
 i{\omega _T}_i^j  &= 
 4d\lambda ^j \varsigma _i  - 4\bar \varsigma ^j d\bar \lambda _i  - 2\bar \varsigma ^j \bar \sigma _\mu  (\lambda ^k \sigma ^\mu  d\bar \lambda _k  - d\lambda ^k \sigma ^\mu  \bar \lambda _k )\varsigma _i  \\ 
 & + 2\bar \varsigma ^j \varsigma _i \sigma _\mu  (\lambda ^k \sigma ^\mu  d\bar \lambda _k  - d\lambda ^k \sigma ^\mu  \bar \lambda _k ) + 2i( - \bar \varsigma ^i \bar \sigma _\mu  \varsigma _j  + \bar \varsigma ^i \varsigma _j \sigma _\mu  )dx^\mu ;  \\ 
 i\omega _M^{ab} & =   
 dx^{\mu '} [2i\phi ^b \delta _{\mu '}^a  + \frac{i}{2}(\bar \varsigma ^i \bar \sigma _{\mu '} \sigma ^{ab} \varsigma _i  - \bar \varsigma ^i \bar \sigma ^{ab} \varsigma _i \sigma _{\mu '} )] - d\lambda ^i \sigma ^{ab} \varsigma _i  + \bar \varsigma ^i \bar \sigma ^{ab} d\bar \lambda _i  \\ 
 & + \frac{1}{2}\bar \varsigma ^i \bar \sigma _\mu  (\lambda ^j \sigma ^\mu  d\bar \lambda _j  - d\lambda ^j \sigma ^\mu  \bar \lambda _j )\sigma ^{ab} \varsigma _i  - \frac{1}{2}\bar \varsigma ^i \bar \sigma ^{ab} \varsigma _i \sigma _\mu  (\lambda ^j \sigma ^\mu  d\bar \lambda _j  - d\lambda ^j \sigma ^\mu  \bar \lambda _j ) \\ 
  &+ 2\phi ^b (\lambda ^i \sigma ^a d\bar \lambda _i  - d\lambda ^i \sigma ^a \bar \lambda _i ); 
\tag{23}
\end{align*}
where the NG fields are the Weyl spinors $\lambda ^i $, $\varsigma _i$ and the axion $a$,  corresponding 
to the broken generators $Q_i ,S^i $ and $A$ respectively.And $\phi^\mu$ is the independent 
collective coordinate in the Coset space. It is not treated as the dynamical field [16,17]. In addition, not all the NG fields are independent 
fields. Note from Eq.(2) that
$$
[\bar S_i^{\dot \alpha } ,P_\mu  ] = \bar \sigma _\mu  ^{\dot \alpha \beta } Q_{\beta i}\\ 
\eqno{(24)}
$$                                                                                                         
Or alternatively, we have
$$
[\bar s_{\dot \alpha }^i \bar S_i^{\dot \alpha }  + s_i^\alpha  S_\alpha ^i ,P_\mu  ] = \bar s_{\dot \alpha }^i \bar 
\sigma _\mu  ^{\dot \alpha \beta } Q_{\beta i}  + s_i^\alpha  \sigma _{\mu
{\alpha \dot \alpha }} \bar Q^{\dot \alpha i}  \\ 
\eqno{(25)}
$$                                                     
Consider the commutator of NG fields $\lambda ^i$  of the broken Supercharges $Q_i $ with Eq.(25),
$$
[[\bar s_{\dot \alpha }^i \bar S_i^{\dot \alpha }  + s_i^\alpha  S_\alpha ^i ,P_\mu  ],\lambda ^{\gamma i} ] = 
[\bar s_{\dot \alpha }^i \bar \sigma _\mu  ^{\dot \alpha \beta } Q_{\beta i}  + s_i^\alpha  
\sigma _{\mu {\alpha \dot \alpha }} \bar Q^{\dot \alpha i} ,\lambda ^{\gamma i} ] \\ 
\eqno{(26)}
$$                                                 
The VEV of the right-hand side can be found from the infinitesimal supersymmetric transformation of $ \lambda ^i$:
$$
< |[\bar s_{\dot \alpha }^i \bar \sigma _\mu  ^{\dot \alpha \beta } Q_{\beta i}  + s_i^\alpha  
\sigma _{\mu {\alpha \dot \alpha }} \bar Q^{\dot \alpha i} ,\lambda ^{\alpha i} ]| > \sim \bar 
s_{\dot \alpha }^i \bar \sigma _\mu  ^{\dot \alpha \alpha }  \\
\eqno{(27)}
$$
                                            
Hence, applying the Jacobin identity to the left-hand side of Eq.(26) yields

$$
< |[\bar s_{\dot \alpha }^i \bar S_i^{\dot \alpha }  + s_i^\alpha  S_\alpha ^i ,i\partial _\mu  \lambda ^{\gamma i} ]| >
 \sim - \bar s_{\dot \alpha }^i \bar \sigma _\mu  ^{\dot \alpha \beta }  \ne 0 \\
\eqno{(28)}
$$                                                                       
On the other hand, the NB fields $\bar \varsigma ^i$ of the broken Supercharge $\bar S_i$ have the infinitesimal supersymmetric 
transformation properties
$$
< |[\bar s_{\dot \alpha }^i \bar S_i^{\dot \alpha }  + s_i^\alpha  S_\alpha ^i ,\bar \varsigma _{\dot \gamma }^i ]| > \sim \bar s_{\dot \gamma }^i  \ne 0 \\
\eqno{(29)}
$$                                                                                       
Therefore, from Eqs.(28,29) we can conclude that
$$
i\partial _\mu  \lambda ^{\gamma i} \sim - \bar \varsigma _{\dot \alpha }^i \bar \sigma _\mu  ^{\dot \alpha \beta }  \\
\eqno{(30)}
$$                                                                                                            
Or it can be re-written as
$$
\bar \varsigma _{\dot \gamma }^i \sim - \frac{1}{4}i\partial _\mu  \lambda ^{\alpha i} \sigma ^\mu  _{\alpha \dot \gamma }  \\
\eqno{(31)}
$$                                                                                                        Consequently, $\bar \varsigma ^i $ are not independent NG fields and can be written as a function of the fields $ \lambda ^i $. 
In order to eliminate them from the effective action, consider the covariant constraints on the one-forms $\omega _Q^i $, i.e., $\omega _Q^i  = 0$. 
Hence, it is found 
$$
\bar \varsigma _{\dot \gamma }^i  =  - \frac{1}{4}ie_a^{ - 1\mu } \partial _\mu  \lambda ^{\alpha i} \sigma ^a _{\alpha \dot \gamma }  \\
\eqno{(32)}
$$
where $e_a^{ - 1\mu }$ is the inverse of the verbein (see Eq.(33) for definition). This is just the inverse Higgs mechanism [4]. Similarly, from $\bar \omega _{\bar Q}  = 0$, it amounts to
$$
\varsigma _i^\alpha   =  - \frac{1}{4}ie_a^{ - 1\mu } \partial _\mu  \bar \lambda _{\dot \gamma i} \bar 
\sigma ^{a {\dot \gamma \alpha }}  \\
\eqno{(32')}
$$
Therefore, the real Nambu-Goldstone (Goldstino) degrees of freedom are these fields $\lambda ^i$ and $a$.  And the 
effective action of these modes can be constructed by using the building blocks of verbein and 
covariant derivatives from the one-forms of Eq.(23).

   Consider the covariant coordinate one-forms $ \omega ^a$, which can be decomposed with respect to the brane world volume  
coordinate differential one-forms $d\xi ^\mu$ as $\omega ^a  =  
 d\xi ^\mu e_\mu ^{\\\ a} $. The verbein is therefore given by
$$
e_\mu ^{\\\ a}  = \frac{\partial x^a}{\partial \xi^\mu}  
-i(\lambda ^i \sigma ^a \frac {\partial \bar \lambda _i}{\partial \xi^\mu}  - \frac {\partial \lambda ^i}{\partial \xi^\mu} \sigma ^a \bar \lambda _i ) \\
\eqno{(33)}
$$                                                                                
where $x^a=(x^0,x^1,x^2,x^3)$, and it becomes
$$
e_\mu ^{\\\ a}  =   
 \delta _\mu ^{\\\ a}  -   
 i(\lambda ^i \sigma ^a \partial _\mu  \bar \lambda _i  - \partial _\mu  \lambda ^i \sigma ^a \bar \lambda _i ) \\
\eqno{(33')}
$$                                                                                in static gauge $\xi ^\mu=x^\mu$. Hence, 
the local Lorentz invariant interval becomes
$$
ds^2  = g_{\mu \nu } d \xi^\mu  d \xi^\nu   = \eta _{ab}  \omega^a  \omega^b  
\\
\eqno{(34)}
$$                                                                                       
where the metric $g_{\mu \nu }  = \eta _{ab} e_\mu ^{\\\ a} e_\nu ^{\\\ b}$. We use Latin letters $a,b$ etc to represent the local tangent coordinate indices, and Greek 
letters $\mu ,\nu $ etc for these of the general coordinates in what follows. We also take the static 
gauge unless explicitly indicated otherwise. In order to construct the covariant 
derivative of the field $a$, since $dx^\mu   = \omega ^a e_a^{ - 1\mu }$, the 
one-forms $\omega _A$ can be re-written as $\omega _A  = \omega ^a D_a a$, in which
\begin{align*}
D_a a &= e_a^{ - 1\mu } 
 (\partial _\mu  a - i2\partial _\mu  \lambda ^i \varsigma _i  + i2\bar \varsigma ^i \partial _\mu  \bar \lambda _i  + i\bar \varsigma ^i \bar \sigma _{\mu '} (\lambda ^j \sigma ^{\mu '} \partial _\mu  \bar \lambda _j  - \partial _\mu  \lambda ^j \sigma ^{\mu '} \bar \lambda _j )\varsigma _i  \\ 
 & - i\bar \varsigma ^i \varsigma _i \sigma _{\mu '} (\lambda ^j \sigma ^{\mu '} \partial _\mu  \bar \lambda _j  - \partial _\mu  \lambda ^j \sigma ^{\mu '} \bar \lambda _j ) + ( - \bar \varsigma ^i \bar \sigma _\mu  \varsigma _i  + \bar \varsigma ^i \varsigma _i \sigma _\mu  )) 
\tag{35}
\end{align*}
On the other hand, under the action of $g$ of Eq.(10), the Maurer-Cartan one-forms transform according to 
$$
\Omega '^{ - 1} d\Omega ' = h(\Omega ^{ - 1} d\Omega )h^{ - 1}  + hdh^{ - 1}  \\
\eqno{(36)}
$$                                                                                     It thus follows that 
the dilatation transformation properties of the verbein and covariant derivative in the local Lorentz 
reference frame are found to be
\begin{align*}
\omega^a &\to e^d \omega^a  \\
e_\mu ^{\hspace{3pt} a}  &\to e^d e_\mu ^{\hspace{3pt} a}  \\  
 D_a a &\to e^{ - d} D_a a 
\tag{37}
\end{align*}
i.e., their scale dimensions are $1,-2 ,$ and $-1$ respectively. In addition, considering the transformation 
of Eq.(37), they are found to be SU(2) singlets.

Introduce an auxiliary (intrinsic) metric $G_{\mu \nu }$ whose scale dimension is $2$ as induced by the scale transformation of $\xi \to e^d \xi^\mu$ on the brane world volume. The local Lorentz invariant interval $ds^2$ has the form 
$$
ds^2  = G_{\mu \nu } d \xi^\mu  d \xi^\nu   \\
\eqno{(38)}
$$ 
which has scale dimension $2$ as a result of scale transformation $ds^2 \to e^{2d} ds^2$.

                                                                                                           Then the 
effective scale invariant action of the brane world volume is given by
\begin{align*}
I_0  &= - \frac{{f_s ^2 }}{2}\int {d^4 \xi} \sqrt {\left| {\det G} \right|} [\frac{1}{4}G^{\mu \nu } \eta _{ab}e^{\hspace{3pt} a}_\mu e^{\hspace{3pt}  b}_\nu]^2   \\
&= - \frac{{f_s ^2 }}{2}\int {d^4 \xi} \sqrt {\left| {\det G} \right|} [\frac{1}{4}G^{\mu \nu } \eta _{ab}   
 (\frac{\partial x^a}{\partial \xi^\mu}  
-i(\lambda ^i \sigma ^a \frac {\partial \bar \lambda _i}{\partial \xi^\mu}  - \frac {\partial \lambda ^i}{\partial \xi^\mu} \sigma ^a \bar \lambda _i ) )\\
  & \cdot (\frac{\partial x^b}{\partial \xi^\nu}  
-i(\lambda ^i \sigma ^b \frac {\partial \bar \lambda _i}{\partial \xi^\nu}  - \frac {\partial \lambda ^i}{\partial \xi^\nu} \sigma ^b \bar \lambda _i ))]^2 
\tag{39}
\end{align*}
in which $\det G = \det G_{\mu \nu }$ and the tensor $G^{\mu \nu }$ with scale dimension is $-2$ is the 
inverse of $G_{\mu \nu } $. The part inside the square brackets has a scale dimension $-2$. It can 
be concluded that Eq.(39) is Weyl scale invariant under the transformation of $\xi^\mu \to e^d \xi^\mu$. 
Obviouely, when the spinors $\lambda^i$ are set to zero, it reduces to the Weyl scale invariant bosonic action [18]:
$$
I_B  = - \frac{{f_s ^2 }}{2}\int {d^4 \xi} \sqrt {\left| {\det G} \right|} [\frac{1}{4}G^{\mu \nu } \eta _{ab}\frac {\partial x^a}{\partial \xi^\mu}\frac {\partial x^b}{\partial \xi^\nu}]^2   
\eqno{(40)}
$$
In static gauge, Eq.(39) becomes
\begin{align*}
I_0  &=  - \frac{{f_s ^2 }}{2}\int {d^4 x} \sqrt {\left| {\det G} \right|} [\frac{1}{4}G^{\mu \nu } \eta _{ab}   
 (\delta _\mu ^{\hspace{3pt} a}  -  
 i(\lambda ^i \sigma ^a \partial _\mu  \bar \lambda _i  - \partial _\mu  \lambda ^i \sigma ^a \bar \lambda _i )) \\
  & \cdot (\delta _\nu ^{\hspace{3pt} b}  -   
 i(\lambda ^i \sigma ^b \partial _\nu  \bar \lambda _i  - \partial _\nu  \lambda ^i \sigma ^b \bar \lambda _i ))]^2  
\tag{41}
\end{align*}
Besides, the spinors $\lambda ^i $ transform as a SU(2) doublet, i.e.
$$
\lambda '^i  = e^{(it_{k'}^{j'} T_{j'}^{k'} )_j^i } \lambda ^j  \\
\eqno{(42)}
$$                                                                                                                
where the SU(2) operators $T_{j'}^{i'}  $, with properties $(T_j^i )^ \dag   = T_i^j $ , have the matrix representation $(T_{j'}^{i'} )_j^i  = \delta _{j'}^i \delta _j^{i'}  - \frac{1}{2}\delta _{j'}^{i'} \delta _j^i  $. Besides, the spinors $\bar \lambda _i $
can be found transforming as SU(2) covariant vectors:
$$
\bar \lambda '_i  = e^{ - (it_{k'}^{j'} T_{j'}^{k'} )_i^j } \lambda _j  \\
\eqno{(43)}
$$                                                                                                              
It is thus obvious that $i(\lambda ^i \sigma ^b \partial _\nu  \bar \lambda _i  - \partial _\nu  \lambda ^i \sigma ^b \bar \lambda _i )$ 
is SU(2) invariant. 
As a result, the action of Eq.(41) is both scale and SU(2) 
invariant.

    When considering the dynamics of the Goldstinos $ \lambda ^i $ alone, the 
auxiliary field $G_{\mu \nu }$ can be eliminated from 
Eq.(41) after applying its equation of motion 
$$
G_{\mu \nu }  = \Lambda e_\mu ^{\\\ a} e_\nu ^{\\\ b} \eta _{ab}  \\
\eqno{(44)}
$$                                                                                                             
where $ \Lambda $ is an arbitrary constant. Then after plugging 
Eq.(44) into Eq.(41), it reduces to
$$
I_0 ' =  - \frac{{f_s ^2 }}{2}\int {d^4 x} \det e_\mu ^{\\\ a}  \\
\eqno{(45)}
$$                                                                                                    
in which the coefficient $f_s$  is related to the SUSY broken scale. It is just the 
Akulov-Volkov action for the case of the extended supersymmetries. Similarly, applying the equation of 
motion of $G_{\mu \nu}$ to the Weyl scale invariant bosonic action of Eq.(40) will lead to the normal Nambu-Goto action
$$
I_B ' = - \frac{{f_s ^2 }}{2}\int {d^4 \xi} \sqrt {\left| \det({\eta _{ab}\frac {\partial x^a}{\partial \xi^\mu}\frac {\partial x^b}{\partial \xi^\nu}} )\right|}  
\eqno{(46)}
$$

    As for the effective scale invariant action of the SU(2) singlet, i.e. the NG field $a$,  it can be constructed by 
using the covariant derivative of Eq.(35):
\begin{align*}
I_1  &=T_A \int {d^4 } x\sqrt {\left |\det G \right |} F(G^{\mu \nu } ,e_\mu^{\hspace{3pt} a})\eta ^{ab} D_a aD_b a 
\tag{47}
\end{align*}
where the compensator function $F(G^{\mu \nu } ,e_\mu ^{\hspace{3pt} a} )$ has the scale dimension $-2$, and 
the coefficient and $T_A $ is related to 
breaking scale of $A$ symmetry. Hence, the complete effective action of 
the fields $\lambda ^i$ and $a$ is given by
\begin{align*}
I &= I_0  + I_1  \\ 
  &= \int {d^4 x} T_A \sqrt {\left | \det G \right |} F(G^{\mu \nu } ,e_\mu ^{\hspace{3pt}a} )\eta ^{ab} D_a aD_b a -  \\ 
 & \frac{{f_s ^2 }}{2}\sqrt {\left| {\det G} \right|} [\frac{1}{4}G^{\mu \nu }   
 (\delta _{\mu \nu }  -   
 i(\lambda ^i \sigma _\mu  \partial _\nu  \bar \lambda _i  - \partial _\nu  \lambda ^i \sigma _\mu  \bar \lambda _i ) \\ 
  & - i(\lambda ^i \sigma _\nu  \partial _\mu  \bar \lambda _i  - \partial _\mu  \lambda ^i \sigma _\nu  \bar \lambda _i ) 
  - (\lambda ^i \sigma _b \partial _\mu  \bar \lambda _i  - \partial _\mu  \lambda ^i \sigma _b \bar \lambda _i ) 
 (\lambda ^i \sigma ^b \partial _\nu  \bar \lambda _i  - \partial _\nu  \lambda ^i \sigma ^b \bar \lambda _i ))]^2  
\tag{48}
\end{align*}    
          
\vspace{5pt}
\begin{flushleft}
{\large IV. Coupling to Matter Fields}
\end{flushleft}
\vspace{5pt}

     Consider the presence of matter fields localized on the brane world volume. They actually behave as the spectator fields, 
transforming covariantly under the unbroken subgroup $H$. Their coupling to those NG fields can 
be obtained by introducing covariant derivatives of the matter fields. Consider matter fields $\Phi _A $, 
in which $A$ represents any internal or Lorentz index. It has a N-dimensional representation of the 
group SU(2), i.e.
$$
\Phi _A  \to \Phi '_A  = e^{it_j^i (T_i^j )_A^B } \Phi _B  \\
\eqno{(49)}
$$                                                                                                  
likewise, its Hermitian conjugate transforms as 
$$
\Phi ^A  \to \Phi '^A  = e^{ - it_i^j (T_j^i )_B^A } \Phi ^B  \\
\eqno{(50)}
$$                                                                                                 
And under the full unbroken subgroup $H$, the matter field would transform according to
$$
\Phi  \to \Phi ' = {\rm{  }}e^{i(fD + m^{\mu \nu } M_{\mu \nu }  + t_j ^i T_i^j )} {\rm{ }}\Phi  \\ 
\eqno{(51)}
$$                                                                                                 
Accordingly, its covariant derivative is given by 
$$
D_\mu  \Phi  = (\partial _\mu   + i\frac{1}{2}\omega _\mu ^{ab} \sum _{ab}  + id_\Phi  \omega _{D_\mu  }^{}  + i\omega _{T\mu j}^i T_i^j )\Phi  \\
\eqno{(52)}
$$                                                                                      
Where ${\sum}_{ab}$ is the representation of the operators $M_{ab}$, and $d_\Phi$ is the scale dimension of the matter field.
The connections $\omega _\mu ^{ab}$, $\omega _{D\mu }$ and $\omega _{T\mu i}^j$ are 
given by the one-forms $\omega _M^{ab}  = dx^\mu  \omega _\mu ^{ab}$,
$\omega _D  = dx^\mu  \omega _{D\mu }$ and $\omega _{Ti}^j  = dx^\mu  \omega _{T\mu i}^j$ respectively. They 
transform inhomogeneously according to Eq.(36). Explicitly, the transformation of the spin connection $\omega _\mu ^{ab}$ 
\begin{align*}
\omega _\mu ^{ab}  &=   
 2\phi ^b \delta _\mu ^{\hspace{3pt}a}  + \frac{1}{2}(\bar \varsigma ^i \bar \sigma _\mu  \sigma ^{ab} \varsigma _i  - \bar \varsigma ^i \bar \sigma ^{ab} \varsigma _i \sigma _\mu  ) + i(\partial _\mu  \lambda ^i \sigma ^{ab} \varsigma _i  - \bar \varsigma ^i \bar \sigma ^{ab} \partial _\mu  \bar \lambda _i ) \\ 
 & - i\frac{1}{2}\bar \varsigma ^i \bar \sigma _{\mu '} (\lambda ^j \sigma ^{\mu '} \partial _\mu  \bar \lambda _j  - \partial _\mu  \lambda ^j \sigma ^{\mu '} \bar \lambda _j )\sigma ^{ab} \varsigma _i  +  \\ 
& i\frac{1}{2}\bar \varsigma ^i \bar \sigma ^{ab} \varsigma _i \sigma _{\mu '} (\lambda ^j \sigma ^{\mu '} \partial _\mu  \bar \lambda _j  - \partial _\mu  \lambda ^j \sigma ^{\mu '} \bar \lambda _j ) 
  - i2\phi ^b (\lambda ^i \sigma ^a \partial _\mu  \bar \lambda _i  - \partial _\mu  \lambda ^i \sigma ^a \bar \lambda _i ) 
\tag{53}
\end{align*} 
is found to have the synthesis form of the general coordinate transformation of Eq.(20) and the inhomogeneous transformation under $h$ according to Eq.(36): 
$$
\omega _{\mu '}^{'cd}  = \frac{{dx^\mu  }}{{dx^{\mu '} }}(\omega _\mu ^{a'b'} \Lambda _{\\\ a'} ^c \Lambda _{\\\ b'} ^d  - 
\partial _\mu  m^{cd} ) \\
\eqno{(54)}
$$                                                                                
where the $\Lambda^c_{\\\ d} $ matrix is defined by the ordinary Lorentz transformation,$ x^{a'}  = \Lambda ^{a'} _{\\\ b} x^b $; and the inhomogeneous 
term can be derived from that of Eq.(36)as following                           
\begin{align*}
hdh^{-1} &=   
 e^{i(fD + m^{\mu \nu } M_{\mu \nu }  + t_j ^i T_i^j )} de^{ - i(fD + m^{\mu \nu } M_{\mu \nu }  + t_j ^i T_i^j )}  \\ 
 & =  - idm^{ab} M_{ab}  - idfD - idt_j^i T_i^j  
\tag{55}
\end{align*} 
which has been expanded up to the leading order of infinitesimal transformation of $g$ in Eq.(9).

     Similarly, it follows from Eq.(23) that the connections $\omega _D  = dx^\mu  \omega _{D\mu }$ has the form
\begin{align*}
\omega _{D\mu }  =  &- 2\phi _\mu   - i(\bar \varsigma ^i \bar \sigma _\mu  \varsigma _i  + \bar \varsigma ^i \varsigma _i \sigma _\mu  ) + 2\partial _\mu  \lambda ^i \varsigma _i  + 2\bar \varsigma ^i \partial _\mu  \bar \lambda _i  \\ 
  &- \bar \varsigma ^i \varsigma _i \sigma _\nu  (\lambda ^j \sigma ^\nu  \partial _\mu  \bar \lambda _j  - \partial _\mu  \lambda ^j \sigma ^\nu  \bar \lambda _j ) + i2\phi ^\nu  (\lambda ^j \sigma _\nu  \partial _\mu  \bar \lambda _j  - \partial _\mu  \lambda ^j \sigma _\nu  \bar \lambda _j ) \\ 
  &- \bar \varsigma ^i \bar \sigma _\nu  (\lambda ^j \sigma ^\nu  \partial _\mu  \bar \lambda _j  - \partial _\mu  \lambda ^j \sigma ^\nu  \bar \lambda _j )\varsigma _i 
\tag{56}
\end{align*} 
in which the non dynamical field $\phi^\mu$ can be eliminated by imposing the covariant constraint on the covariant curl $C_{\mu \nu }^a$  [17], i.e.
\begin{align*}
C_{\mu \nu }^a  = 0 &= D_\mu  e_\nu  ^{\hspace{3pt} a}  - D_\nu  e_\mu  ^{\hspace{3pt} a}  \\ 
  &= \partial _\mu  e_\nu  ^{\hspace{3pt} a}  + i\frac{1}{2}\omega _\mu ^{a'b'} (\sum _{a'b'} )^a _{\hspace{3pt} b} e_\nu  ^{\hspace{3pt} b}  + 
i\omega _{D\mu }^{} e_\nu  ^{\hspace{3pt} a}  - \partial _\nu  e_\mu  ^{\hspace{3pt} a}  \\ 
  &- i\frac{1}{2}\omega _\nu ^{a'b'} (\sum _{a'b'} )^a _{\hspace{3pt} b} e_\mu  ^{\hspace{3pt} b}  - i\omega _{D\nu }^{} e_\mu  ^{\hspace{3pt} a}  
\tag{57}
\end{align*}

    Consider the scalar field $\phi$, which has trivial representation of $M_{ab}$. Therefore, the covariant 
derivative becomes $D_\mu  \phi  = (\partial _\mu   - i\omega _{D\mu }  + i\omega _{T\mu j}^i T_i^j )\phi$. Then the scale invariant action of the scalar field is shown to be
\begin{align*}
I_\Phi  & = \int {d^4 } x\det eg^{\mu \nu } D_\mu  \phi D_\nu  \phi \\ 
 &= \int {d^4 } x\det e\eta ^{ab} e_a^{ - 1\mu } e_b^{ - 1\nu } (\partial _\mu   - i\omega _{D\mu }  + i\omega _{T\mu j}^i T_i^j )\phi  \\ 
  &\cdot (\partial _\nu   + i\omega _{D\nu }  - i\omega _{T\nu j}^i T_i^j )\phi  
\tag{58}
\end{align*}                                                 
and the scalar field has scale dimension $-1$ . Considering Eq.(49) and (50), it is also obviously 
SU(2) invariant.

     Similarly, consider the fermion fields $\psi$, which give a nontrivial representation of $M_{ab} $, i.e. $ {\sum} _{ab}  = \frac{1}{2}i(\gamma _a \gamma _b  - \gamma _b \gamma _a )$ . 
Its covariant derivative is then found to be 
$ D_\mu  \psi  = (\partial _\mu   + i\frac{1}{2}\omega _\mu ^{ab} {\sum} _{ab}  - i\frac{3}{2}\omega _{D\mu }  + i\omega _{T\mu j}^i T_i^j )\psi$. 
Thus its scale and SU(2) invariant action is given by
\begin{align*}
I_\psi   = & \int {d^4 } x\det e\bar \psi i\gamma ^\mu  D_\mu  \psi  \\
 & = \int {d^4 } x\det e\bar \psi i\gamma ^a e_a^{ - 1\mu } (\partial _\mu   + i\frac{1}{2}\omega _\mu ^{ab} {\sum} _{ab}  - i\frac{3}{2}\omega _{D\mu }  + i\omega _{T\mu j}^i T_i^j )\psi 
\tag{59}
\end{align*}                                                                                             
in which the fermion field has the scale dimension $ - \frac{3}{2}$. Therefore the total action of 
the matter fields coupling to these NG fields can be described by
\begin{align*}
I_M & = I_\Phi   + I_\Psi   \\ 
&= \int {d^4 } x\det e(\eta ^{ab} e_a^{ - 1\mu } e_b^{ - 1\nu } (\partial _\mu   - i\omega _{D\mu }  + i\omega _{T\mu j}^i T_i^j )\phi  \\ 
 & \cdot (\partial _\nu   + i\omega _{D\nu }  - i\omega _{T\nu j}^i T_i^j )\phi  \\ 
 & + \bar \psi i\gamma ^a e_a^{ - 1\mu } (\partial _\mu   + i\frac{1}{2}\omega _\mu ^{ab} {\sum} _{ab}  - i\frac{3}{2}\omega _{D\mu }  + i\omega _{T\mu j}^i T_i^j )\psi ) 
\tag{60}
\end{align*}

 As a result, on the brane world volume the unbroken symmetries $H$ are realized linearly on the localized spectator fields, 
whileas the broken symmetries are realized nonlinearly through the Nambu-Goldstone(Goldstino) modes $a$ and $\lambda^i$. On the 
other hand, the full symmetry can also be directly realized on the field itself, such as the standard realization [15, 19]. One 
can start from the linear transformation of the matter fields under the stability group $H$, then promote it to describe the full 
symmetry by taking the parameters of $h$ as these induced form the left action of $G$ on the Coset representatives through 
Eq.(10). Hence the transformation of the fields can realize the full symmetry group $G$, but nonlinearly.

    The Weyl scale invariant brane dynamics has also been extensively studied in [18,20-23]. The action of Eq.(48), being 
a low energy effective theory, describes the long wave oscillations of the brane into the Grassmann coordinates of the 
superspace along with the effective dynamics of the localized Nambu-Goldstone mode $a$ corresponding to the broken $A$ symmetry. 
Consequently, by using the approach of nonlinear realization, the total action of Eq.(48) gives us an effective theory describing the 
spontaneously breaking of a larger symmetry group, i.e. the N=2 superconformal symmetry. In fact, the Weyl scale invariant brane 
action Eq.(41) is an extension of its bosonic counterpart action Eq.(40) to include the fermionic sectors as a result of the full 
supersymmetry breaking. Therefore, the fields localized on the brane world volume would be described by the Weyl scale symmetry theory [13, 24-27].

    Besides, the internal symmetry $SU(2) \times U(1)_R $ of the extended N=2 Superconformal symmetry can also 
be broken down to $U(1)_R $[28]. Hence, corresponding to generators $T_i^j $, there would be three NG scalar 
fields present in the effective action along with other possible $SU(2)$ broken terms. 
The existence of these NG particles would give rise to long range forces in nature, 
which may affect astrophysical considerations such as contributing new mechanisms for 
energy loss from stars [19].
      
    In addition, it is noteworthy that if one takes the Lorentz group as the stability group instead, the dilatation symmetry is also spontaneously broken. Therefore, 
it will result an additional NG field, the dilaton $\sigma$ , whose intrinsic transformation yields 
$$ 
\sigma '\sim d + \sigma
\eqno{(61)}                                                                                         $$                       
Therefore, the scale invariant effective action of the Goldstino fields becomes 
$$
I_0  =  - T\int {d^4 x} e^{ - 4\sigma } \det e \\
\eqno{(62)}                                                                                         
$$ 
where the dilaton is introduced as the compensator field. From Eq.(33`), it can be concluded that there is a constant term in det$e$  which 
shifts the VEV and signals the spontaneous breaking of the SUSY. When considering together with the compensator field $ e^{ - 4\sigma }$, the VEV 
can be determined by estimating the value of $< |e^{ - 4\sigma } | >$, which becomes minimum when $ < |\sigma | > $  goes to $\propto $. 
Therefore, due to the unbound 
of the VEV of the dilaton field it follows that $\sigma $ can not be a NG particle. It thus indicates the incompatibility of the nonlinear realization of 
the SUSY and the dilatation symmetry when taking dilaton as the compensator field [10, 29]. It is interesting to note that if one takes the 
Lorentz group as the stability group and works on the superspace parameterized 
by the Coset space, it will lead to the supersymmetric theory (superbrane) as discussed in [5, 30]. However, this is not our case in the 
present context. 
                                                                                                  
Besides, due to the absence of the dilaton in the effective action (41), according to the $AdS/CFT $correspondence, 
one may expect it is feasible to embed a probe brane in a supersymmetric $AdS \times S$ space. And following the 
outline of the present paper, the supersymmetric isometry group of the background space can be realized through
 the dynamics of the brane but with no destabilized terms resulting from oscillations in the transverse spatial directions.
 That would be instructive to explore more aspects about $AdS/CFT $ correspondence. 
Further work about this correspondence is being investigated and it would be of interest to the theory of brane world scenarios as well.

$$
\\
$$

The author thanks the support of the HEP group of Physics Department at Purdue
University. The author also thanks Muneto Nitta for some useful comments.

\pagebreak

\begin{flushleft}
{\large Appendix }
\end{flushleft}
\vspace{5pt}

The $A$s and $B$s in Eqs.(20,21,23) are defined as following:\\ 
(i)
\begin{align*}
e^{ - i(\varsigma S + \bar \varsigma \bar S)} a^\mu  P_\mu  e^{i(\varsigma S + \varsigma \bar S)}  
=\ldots +\frac{1}{{3!}}(A_{1k} S^k  + B_1^k \bar S_k )+ \ldots 
  \tag{{\it{A}}.1}
\end{align*}
where
\begin{align*}
A_{1k}  &=  - i(\bar \varsigma ^i \bar \sigma _{\mu '} a^{\mu '} \sigma ^{\mu \nu } \varsigma _i  - \bar \varsigma ^i \bar \sigma ^{\mu \nu } \varsigma _i \sigma _{\mu '} a^{\mu '} )\frac{1}{2}\varsigma _k \sigma _{\mu \nu }  \\
&+ 4i( - \bar \varsigma ^i \bar \sigma _\mu  a^\mu  \varsigma _j  + \bar \varsigma ^i \varsigma _j \sigma _\mu  a^\mu  )\varsigma _i \delta _k^j  
+ 2i\bar \varsigma ^i \bar \sigma _\mu  a^\mu  \varsigma _i \varsigma _k  \tag{{\it{A}}.2} \\
  B_1^k  &=  - i(\bar \varsigma ^i \bar \sigma _{\mu '} a^{\mu '} \sigma ^{\mu \nu } \varsigma _i  - \bar \varsigma ^i \bar \sigma ^{\mu \nu } \varsigma _i \sigma _{\mu '} a^{\mu '} )\frac{1}{2}\bar \varsigma ^k \bar \sigma _{\mu \nu }  \\ 
  &- 4i( - \bar \varsigma ^i \bar \sigma _\mu  a^\mu  \varsigma _j  + \bar \varsigma ^i \varsigma _j \sigma _\mu  a^\mu  )\bar \varsigma ^j \delta _i^k   
  + 2i\bar \varsigma ^i \varsigma _i \sigma _\mu  a^\mu  \bar \varsigma ^k  
  \tag{{\it{A}}.3}
\end{align*}
(ii) 
\begin{align*}
e^{ - i(\varsigma S + \bar \varsigma \bar S)} qQe^{i(\varsigma S + \varsigma \bar S)} = \ldots +
 \frac{1}{2}(A_{2k} S^k  + B_2^k \bar S_k ) + \ldots 
\tag{{\it{A}}.4}
\end{align*}
where
\begin{align*}
A_{2k}  &= \frac{1}{2}q^i \sigma ^{\mu \nu } \varsigma _i \varsigma _k \sigma _{\mu \nu }  + 4q^i \varsigma _j (\varsigma _i \delta _k^j  - \frac{1}{2}\delta _i^j \varsigma _k )  
  \tag{{\it{A}}.5} \\
B_2^k  &= \frac{1}{2}q^i \sigma ^{\mu \nu } \varsigma _i \bar \varsigma ^k \bar \sigma _{\mu \nu }  - 4q^i \varsigma _j \bar \varsigma ^j \delta _i^k 
  \tag{{\it{A}}.6}
\end{align*}
(iii)   
\begin{align*}
e^{ - i(\varsigma S + \bar \varsigma \bar S)} \bar q\bar Qe^{i(\varsigma S + \varsigma \bar S)} = \ldots +
 \frac{1}{2}(A_{3k} S^k  + B_3^k \bar S_k )+ \ldots 
\tag{{\it{A}}.7}
\end{align*}

where
\begin{align*}
A_{3k}  &=  - \frac{1}{2}\bar \varsigma ^i \bar \sigma ^{\mu \nu } \bar q_i \varsigma _k \sigma _{\mu \nu }  - 4\bar \varsigma ^j \bar q_i \varsigma _j \delta _k^i  
 \tag{{\it{A}}.8} \\
B_3^k  &=  - \frac{1}{2}\bar \varsigma ^i \bar \sigma ^{\mu \nu } \bar q_i \bar \varsigma ^k \bar \sigma _{\mu \nu }  + 4\bar \varsigma ^j \bar q_i (\bar \varsigma ^i \delta _j^k  - \frac{1}{2}\delta _j^i \bar \varsigma ^k ) 
  \tag{{\it{A}}.9}
\end{align*}
(iv)   
\begin{align*}
e^{ - i(\lambda Q + \bar \lambda \bar Q)} sSe^{i(\lambda Q + \bar \lambda \bar Q)} = \ldots +\frac{1}{2}(A_4^k Q_k  + B_{4k} \bar Q^k ) + \ldots 
\tag{{\it{A}}.10}
\end{align*}

where
\begin{align*}
A_4^k & =  - \lambda ^i \sigma ^{\mu \nu } s_i \frac{1}{2}\lambda ^k \sigma _{\mu \nu }  + 4\lambda ^i s_j (\lambda ^j \delta _i^k  - \frac{1}{2}\delta _i^j \lambda ^k )  \tag{{\it{A}}.11}
\\ 
 B_{4k}  &=  - \lambda ^i \sigma ^{\mu \nu } s_i \frac{1}{2}\bar \lambda _k \bar \sigma _{\mu \nu }  - 4\lambda ^i s_j \bar \lambda _i \delta _k^j  \tag{{\it{A}}.12}
\end{align*}
(v)               
\begin{align*}
e^{ - i(\lambda Q + \bar \lambda \bar Q)} \bar s\bar Se^{i(\lambda Q + \bar \lambda \bar Q)}  =\ldots +
 \frac{1}{2}(A_5^k Q_k  + B_{5k} \bar Q^k ) + \ldots 
\tag{{\it{A}}.13}
\end{align*}
where
\begin{align*}
A_5^k  &= \bar s^i \bar \sigma ^{\mu \nu } \bar \lambda _i \frac{1}{2}\lambda ^k \sigma _{\mu \nu }  - 4\bar s^i \bar \lambda _j \lambda ^j \delta _i^k  \tag{{\it{A}}.14}\\ 
 B_{5k}  &= \bar s^i \bar \sigma ^{\mu \nu } \bar \lambda _i \frac{1}{2}\bar \lambda _k \bar \sigma _{\mu \nu }  + 4\bar s^i \bar \lambda _j (\bar \lambda _i \delta _k^j  - \frac{1}{2}\delta _i^j \bar \lambda _k ) 
  \tag{{\it{A}}.15}
\end{align*} 
(vi)             
\begin{align*}
e^{ - i(\lambda Q + \bar \lambda \bar Q)} b^\mu  K_\mu  e^{i(\lambda Q + \bar \lambda \bar Q)} = \ldots + \frac{1}{{3!}}(A_6^k Q_k  + B_{6k} \bar Q^k )+ \ldots 
\tag{{\it{A}}.16}
\end{align*}
where
\begin{align*}
A_6^k & = - i[\frac{1}{2}( - \lambda ^i \sigma ^{\mu \nu } b^{\alpha '} \bar \lambda _i \bar \sigma _{\alpha '}  + b^{\mu '} \lambda ^i \sigma _{\mu '} \bar \sigma ^{\mu \nu } \bar \lambda _i )\lambda ^k \sigma _{\mu \nu }  - 2b^\mu  \lambda ^i \bar \lambda _i \bar \sigma _\mu  \lambda ^k  +  \\ 
 &  4b^\mu  (\lambda ^i \bar \lambda _j \bar \sigma _\mu   - \lambda ^i \sigma _\mu  \bar \lambda _j )\lambda ^j \delta _i^k ] 
  \tag{{\it{A}}.17}\\
B_{6k} & =  - i[\frac{1}{2}( - \lambda ^i \sigma ^{\mu \nu } b^{\alpha '} \bar \lambda _i \bar \sigma _{\alpha '}  + b^{\mu '} \lambda ^i \sigma _{\mu '} \bar \sigma ^{\mu \nu } \bar \lambda _i )\bar \lambda _k \bar \sigma _{\mu \nu }  - 2b^\mu  \lambda ^i \sigma _\mu  \bar \lambda _i \bar \lambda _k  -  \\ 
 &  4b^\mu  (\lambda ^i \bar \lambda _j \bar \sigma _\mu   - \lambda ^i \sigma _\mu  \bar \lambda _j )\bar \lambda _i \delta _k^j  
   \tag{{\it{A}}.18}
\end{align*}
(vii)   In Eq.(23), $ X^\mu   $ is defined as following
\begin{align*}
X^\mu   =& 2[ - d\lambda ^i \sigma ^{\mu '\mu } \varsigma _i  + \bar \varsigma ^i \bar \sigma ^{\mu '\mu } d\bar \lambda _i  + \frac{1}{2}(\bar \varsigma ^i \bar \sigma _\nu  a^\nu  \sigma ^{\mu '\mu } \varsigma _i  - \bar \varsigma ^i \bar \sigma ^{\mu '\mu } \varsigma _i \sigma _\nu  a^\nu  )]\phi _{\mu '} \\
&- (2id\lambda ^i \varsigma _i  + 2i\bar \varsigma ^i d\lambda _i  - i\varsigma ^i \bar \sigma _{\mu '} a^{\mu '} \varsigma _i  - i\bar \varsigma ^i \varsigma _i \sigma _{\mu '} a^{\mu '} )\phi ^\mu   
  + \frac{1}{{3!}}2i(B_2^k \bar \sigma ^\mu  \varsigma _k  - \bar \varsigma ^k \bar \sigma ^\mu  A_{2k} ) +  \\ 
&  + \frac{1}{{3!}}2i(B_3^k \bar \sigma ^\mu  \varsigma _k  - \bar \varsigma ^k \bar \sigma ^\mu  A_{3k} ) +  
 \frac{1}{{4!}}2i(B_1^k \bar \sigma ^\mu  \varsigma _k  - \bar \varsigma ^k \bar \sigma ^\mu  A_{1k} ) - \phi ^{\mu '} \phi _{\mu '} a^\mu   
\tag{{\it{A}}.19}
\end{align*}
where $a^\mu   = \lambda \sigma ^\mu  d\bar \lambda  - d\lambda \sigma ^\mu  \bar \lambda $.

\pagebreak

\begin{center}
{\bf REFERENCES}
\end{center}
\begin{description}

\item[[1]]  M.Gell-Mann and M.Levy, Nuovo Cim.16:705,1960; S.Weinberg,Phys.Rev. 166, 
     1568, 1968; W.A.Bardeen and B.W. Lee, Phys.Rev.177:2389, 1969.

\item[[2]] S. Coleman, J.Wess, and B. Zumino, Phys. Rev. 177, 2239 (1969); C. Callan, S. 
     Coleman, J.Wess, and B. Zumino, Phys. Rev. 177, 2247 (1969).
\item[[3]] C.J. Isham, Abdus Salam, J.A. Strathdee, Annals Phys.62:98-119, 1971; A.B.Borisov, V.I. Ogievetsky, Theor.Math.Phys.21:1179, 1975, Teor.Mat.Fiz.21:329- 
     342, 1974
\item[[4]] E.A. Ivanov and V.I. Ogievetsky, Teor.Mat.Fiz.25:164-177, 1975
\item[[5]] P. West, JHEP 0002, 24(2000)
\item[[6]] For example, see J. Hughes and J. Polchinski, Nucl. Phys. B278, 147 (1986);
E. Ivanov and S. Krivonos, Phys. Lett. B 453, 237 (1999).Erratum ibid. B657: 269,
2007; C. P. Burgess, E. Filotas, M. Klein, and F. Quevedo, J. High Energy Phys.0310
(2003) 41; F. Gonzalez-Rey, L.Y. Park, and M. Rocek, Nucl. Phys. B544, 243 (1999);
T.E. Clark, M. Nitta, T. ter Veldhuis, Phys.Rev.D70:125011, 2004; T.E. Clark,
S.T. Love, M. Nitta, T. ter Veldhuis , J.Math.Phys.46:102304,2005
\item[[7]] For example, for non-BPS branes, see T.E. Clark, M. Nitta and T. ter Veldhuis, Phys.Rev.D67: 085026, 2003; Lu-Xin Liu, arXiv: 0711.4868[hep-th]; 
Lu-Xin Liu, arXiv: hep-th/0611212
\item[[8]] For example, see N.Seiberg and E.Witten, Nucl. Phys. B 426, 19 (1994); Nucl. Phys. B 431, 484 (1994) 
\item[[9]] Y.Gotoh, T.Uematsu, Phys.Lett.B420:69-76,1998; K.Kobayashi, K.Lee, T.Uematsu, Nucl.Phys.B309:669, 1988;
\item[[10]] T.E. Clark and S.T. Love, Phys.Rev.D61:057902, 2000
\item[[11]] For a review, see J. Maldacena, arXiv: hep-th/0309246; I. R. Klebanov, arXiv: hep-th/0009139
\item[[12]] S. Bellucci, E. Ivanov and S. Krivonos,Phys.Rev.D66:086001, 2002; Erratum-ibid.D67:049901,2003
\item[[13]] M. Blagojevic, Gravitation and Gauge Symmetries, IOP Publishing, 2002
\item[[14]] D. V. Volkov and V. P. Akulov, JETP Lett. 16, 438(1972); Phys. Lett. B46, 109(1973)    
\item[[15]] J.Wess and J.Bagger, Supersymmetry and Supergravity, Princeton, 1992
\item[[16]] E.A. Ivanov and J. Niederle, Phys.Rev.D25, 988, 1982
\item[[17]] E.A. Ivanov and J. Niederle, Phys.Rev.D25, 976, 1982
\item[[18]] M. S. Alver and J. Barcelos-Neto, Europhys. Lett. 7, 395 (1988); 8, 90(Erratum)(1989)
\item[[19]] Lu-Xin Liu, Mod.Phys.Lett.A20, 2545, 2005
\item[[20]] A.Aurilia, A.Smailagic and E.Spallucci, Phys.Rev.D51:4410, 1995
\item[[21]] J.Antonio Garcia, Roman Linares and J.David Vergara, Phys. Lett. B503, 154(2001)
\item[[22]] C.Alvear, R.Amorim and J.Barcelos-Neto, Phys. Lett. B273, 415(1991)
\item[[23]] Lu-Xin Liu, Phys. Rev. D74, 45030(2006)
\item[[24]] Ryoyu Utiyama, Progress of Theor. Phys.50,  2080, 1973; 
Progress of Theor.Phys.53, 565, 1975
\item[[25]] Kenji Hayashi and T.Kugo, Progress of Theor. Phys.61, 334, 1979
\item[[26]] A. Bregman, Progress of Theor. Phys.49, 667, 1973
\item[[27]] Kenji Hayashi, M.Kasuya and T.Shirafuji, Progress of Theor. Phys.57, 431, 1977
\item[[28]] P.Fayet, Nucl. Phys. B149, 137, 1979
\item[[29]] T.E. Clark and S.T. Love, arXiv: hep-th/0510274
\item[[30]] S.M. Kuzenko, Ian N. McArthur, Phys.Lett.B522:320, 2001; F. Delduc, E. Ivanov, S. Krivonos, Phys.Lett.B529:233, 2002; P. Pasti , 
D.P. Sorokin, M. Tonin, arXiv:hep-th/9912076

\end{description}

\end{document}